\documentclass[12pt]{article}
\usepackage{aasms}

\begin{document}

\title{Does Pulsar B1757--24 Have a Fallback Disk?}

\author{D. Marsden\footnote[1]{NAS/NRC Research Associate}}
\affil{NASA/Goddard Space Flight Center, Code 662, Greenbelt, MD 20771}

\author{R. E. Lingenfelter and R. E. Rothschild}
\affil{Center for Astrophysics and Space Sciences, University of 
California at San Diego \\ La Jolla, CA 92093}

\begin{abstract}

Radio pulsars are thought to spin-down primarily due to torque 
from magnetic dipole radiation (MDR) emitted by the time-varying 
stellar magnetic field as the star rotates. This assumption 
yields a `characteristic age' for a pulsar which has generally 
been assumed to be comparable to the actual age. Recent 
observational limits on the proper motion of pulsar B1757--24, 
however, revealed that the actual age ($>39$ kyr) of this pulsar 
is much greater than its MDR characteristic age ($16$ kyr) -- 
calling into question the assumption of pure MDR spin-down for 
this and other pulsars. To explore the possible cause of this 
discrepancy, we consider a scenario in which the pulsar acquired 
an accretion disk from supernova ejecta, and the subsequent spin-down 
occurred under the combined action of MDR and accretion torques. A 
simplified model of the accretion torque involving a constant mass 
inflow rate at the pulsar magnetosphere can explain the age and 
period derivative of the pulsar for reasonable values of the pulsar 
magnetic field and inflow rate. We discuss testable predictions of 
this model. 

\end{abstract} 


\section{Introduction}
\label{intro}

Isolated pulsars are spinning neutron stars whose observed spin 
rates gradually decrease with time. The age $\tau$ of a pulsar 
is usually assumed to be equal to the timing (or characteristic) 
age $\tau_{MDR}$ derived by assuming pure magnetic dipole 
spin-down {\it in vacuo}, and the age is then given by $\tau_{MDR}=
-\Omega/(2\dot{\Omega})$ (e.g. \cite{manchester77}), where 
$\Omega=2\pi/P$ and $\dot{\Omega}=-2\pi\dot{P}/P^{2}$ are the 
angular spin frequency and angular frequency derivative for a 
pulsar spin period $P$ and period derivative $\dot{P}$. Under 
this same assumption of MDR spin-down, the magnetic field 
strength of the pulsar is given by the formula (\cite{manchester77})
\begin{equation}
\label{eq0}
B=3.2\times10^{19}{(P\dot{P})}^{1/2}\rm{G}, 
\end{equation}
which is often assumed to be equal to the true field strength of 
the isolated pulsar (e.g. \cite{taylor93}).

PSR B1757--24 is a $0.125$ s radio pulsar which appears to be 
associated with the supernova remnant G5.4--1.2 (\cite{caswell87}). 
The pulsar is surrounded by a compact radio nebula having a cometary 
morphology with a tail extending back into the supernova remnant 
(\cite{frail94}) -- strongly suggesting that the pulsar was formed 
in the supernova which produced G5.4--1.2 (\cite{manchester91}). 
Given the temporal parameters of PSR B1757--24 ($P=0.125$ s and 
$\dot{P}=1.28\times 10^{-13}$ s s$^{-1}$; \cite{taylor93}), 
$\tau_{MDR}=16$ kyr and $B=4\times 10^{12}$ G for the pulsar. 
Assuming that $\tau_{MDR}$ is similar to the pulsar's its true 
age, the transverse velocity implied by the pulsar's displacement 
from the apparent center of the G5.4--1.2 is greater than $1500$ 
km s$^{-1}$ (\cite{gaensler00}; \cite{frail91}). Observations of 
PSR B1757--24 taken six years apart, however, failed to detect the 
expected proper motion from the pulsar, yielding a distance-independent 
lower limit on the age of PSR B1757--24/G5.4--1.2 of $39$ kyr 
(\cite{gaensler00}). This is more than a factor of two greater than 
the pulsar's MDR characteristic age.

The discrepancy between the proper motion age and the MDR characteristic 
age of PSR B1757--24 suggests that the spin-down of the pulsar is 
not due purely to MDR but also has significant contributions from other 
sources of torque. Istomin (1994) considered a model for PSR B1757--24 
in which the pulsar was interacting with dense plasma in the shell of 
G5.4--1.2, causing an   increase in the torque at the light cylinder. 
Here we consider another possible source of extra torque on the neutron 
star -- from a disk of material accreted from ejecta produced in the 
supernova explosion. These {\it fallback disks} may be roughly divided 
into two categories: ``prompt'' and ``delayed''. Prompt disks may be 
formed from $\sim 0.001-0.1M_{\odot}$ (\cite{michel88}; \cite{lin91}) 
of ejecta material soon after the initial core collapse in a type II 
supernova explosion (\cite{woosley95}). Formation of such prompt disks 
is probably limited to $<7$ days after the core collapse because of 
heating of the ejecta by $^{56}$Ni decays (\cite{chevalier89}). Delayed 
disks may form years after the explosion from ejecta decelerated by 
radiative cooling (\cite{fryer99}) or by a strong reverse shock 
(\cite{truelove99}) caused by the primary supernova blast wave 
impinging on dense circumstellar material from the pre-supernova 
stellar wind (\cite{gaensler99}). Whether or not a neutron star 
accretion disk can form shortly after a supernova explosion depends 
on the opposing forces of the pulsar MDR wind and the pressure of 
the hot and turbulent environment shortly after the explosion. 
Since the latter is highly uncertain (\cite{woosley95}), for the 
purposes of this {\it Letter} we assume that a disk can form around 
a neutron star under these conditions and explore the implications 
for PSR B1757--24. 

\section{Spin-down From Accretion Torques}

An accretion disk around a magnetized neutron star can exert a spin-down 
torque on the star if the mass inflow rate is low and the magnetic field 
is strong. Quantitatively, this condition is met when the Keplerian 
co-rotation radius $R_{c}=1.7\times 10^{8}P^{2/3}$ cm is less than the 
magnetospheric radius $R_{M}=4.6\times 10^{8}B_{12}^{4/7}\dot{m}_{15}^{-2/7}$ 
cm, where $\dot{m}=10^{15}\dot{m}_{15}$ g s$^{-1}$ is the mass inflow rate 
onto the magnetosphere and $B=10^{12}B_{12}$ G is the strength of the 
neutron star dipole magnetic field (see e.g \cite{frank92}). Here and 
elsewhere we assume a neutron star mass and radius of $M_{\ast}=1.4
M_{\odot}$ and $R_{\ast}=10$ km, respectively. When $R_{M}>R_{c}$, 
infalling material is stopped at the magnetosphere by a centrifugal 
barrier which prevents accretion onto the neutron star surface. 
Instead, the infalling material may be accelerated away in a 
wind which carries away angular momentum from the magnetosphere 
and hence the neutron star itself. This ``propeller effect'' 
(\cite{illarionov75}) spin-down mechanism has been invoked to 
explain the behavior of some Galactic accretion--powered x--ray 
pulsars (\cite{cui97}; \cite{zhang98}), the spin-evolution of 
anomalous x--ray pulsars (AXPs: \cite{vanparadijs95}; \cite{chatterjee00}) 
and soft gamma--ray repeaters (\cite{alpar00}; \cite{marsden00}), and 
the optical and infrared spectra of some radio pulsars (\cite{perna00}). 

The pulsar B1757--24 is in the propeller regime ($R_M>R_C$) for mass 
inflow rates of $\dot{m}<5\times 10^{17}$ g s$^{-1}$ and a canonical 
(\cite{manchester77}; \cite{taylor93}) neutron star magnetic 
field of $\sim 10^{12}$ G. Since the characteristic age for an old 
($>10$ kyr) pulsar depends only weakly on the initial spin period 
(\cite{manchester77}), we neglect the different formation times 
of the delayed and prompt fallback disks, and assume that the 
spin-down evolution of the pulsar at all times is determined by 
the combined torque from both the MDR and the accretion disk. The 
spin-down rate due to radiation from a rotating magnetic dipole is 
(\cite{manchester77})
\begin{equation}
\label{mdr}
\dot{\Omega}_{M}=-{2B^{2}R_{\ast}^{6}\Omega^{3}\over 3I_{\ast}c^{3}},
\end{equation}
where $I_{\ast}={2\over 5}M_{\ast}R_{\ast}^{2}$ is the neutron star 
moment of inertia. The spin-down rate due to the propeller torque 
vanishes gradually (\cite{chatterjee00}; \cite{alpar00}) as the star 
approaches spin-equilibrium ($R_{M}=R_{c}$) at a spin period $P_{eq}=
4.7B_{12}^{6/7}{\dot{m}_{15}}^{-3/7}$ s. The spin-down rate due 
to the propeller torque is simply (\cite{menou99})
\begin{equation}
\label{wind}
\dot{\Omega}_{A}=k{\dot{m}R_{M}^{2}\Omega_{eq}\over I_{\ast}}\left(1-
{\Omega\over \Omega_{eq}}\right),
\end{equation}
where $k$ is a positive constant of order unity (\cite{wang85}) and 
$\Omega_{eq}=2\pi/P_{eq}$. Assuming a constant mass inflow rate 
and dipole magnetic field, the timing age for a final spin 
period $P=2\pi/\Omega$ is given by
\begin{equation}
\label{age}
\tau_{comb}=\int_{\Omega_{0}}^{\Omega}{d\Omega\over \dot{\Omega}_{M} + 
\dot{\Omega}_{A}},
\end{equation}
where $\Omega_{0}$ is the initial angular frequency. A more realistic 
expression for the propeller torque would incorporate a time dependent 
mass inflow rate (\cite{cannizzo90}) in Equation (\ref{wind}), as 
$\dot{m}$ should decrease in time as the disk dissipates. For the 
model to be correct, however, a disk must still be present around the 
pulsar, because otherwise the MDR timing age would be greater than 
the true age. Therefore the $\dot{m}$ used here may be thought of 
as a time-averaged value of the mass inflow rate. In addition, the 
effect of the propeller flow on the MDR torque (\cite{roberts73}) 
is not taken into account. We plan on incorporating both of these 
effects in future work.

A contour plot of the pulsar B1757--24 timing age $\tau_{comb}$ for 
various values of the magnetic field strength $B$ and mass inflow 
rate $\dot{m}$ is show in Figure 1. The characteristic ages were 
calculated using Equations (\ref{mdr}--\ref{age}) and assuming 
$P=0.125$ s, $P_{0}=10$ ms, and $k=1$. In this simple model, the 
allowed values of $B$ and $\dot{m}$ for pulsar B1757--24 lie on the 
heavy solid line corresponding to the observed $P$ and $\dot{P}$. 
In addition, the shaded regions of parameter space in Figure 1 are 
excluded by the lower limit on the age (right shaded region: 
\cite{gaensler00}), and the necessary condition $R_{M}>R_{c}$ 
(shaded region in upper left corner). From Figure 1, we find 
that values of $2\times 10^{11}<B<1.4\times 10^{12}$ G, $7\times 
10^{13}<\dot{m}<7\times 10^{17}$ g s$^{-1}$, and $39<\tau_{comb}<60$ 
kyr are consistent with the lower limit on the true age of $39$ kyr 
(\cite{gaensler00}) and the present-day spin-down rate of the pulsar. 

\section{Discussion}

In the context of this model, the required mass inflow rate for 
PSR B1757--24 overlaps the range of $\dot{m}$ inferred from 
accretion--powered neutron star systems (\cite{bildsten97}). 
Radio pulsations are not seen from accretion--powered x--ray 
pulsars in binary systems (\cite{fender96}), which implies 
that the emission mechanism responsible for radio pulsations 
may be quenched by matter near the polar caps of the pulsar. 
This is not a problem for the PSR B1757--24 model, however, 
because in propeller sources most of the matter is ejected before 
it has a chance to reach the polar cap and quench the radio 
emission. Radio emission may be suppressed for propeller sources 
closer to spin equilibrium than PSR B1757--24 (e.g. the AXPs: 
\cite{chatterjee00}), however, because as equilibrium is approached 
more matter will be allowed to fall onto the neutron star surface. 
The total mass required to fuel the propeller spin-down over the 
lifetime (so far) of pulsar B1757--24 would be $\dot{m}\tau\sim 
10^{-8}-10^{-3}M_{\odot}$ -- a tiny fraction of the total amount 
of ejecta in a typical Type II supernova explosion (\cite{woosley95}).

This hypothesis can be tested by multiwavelength observations. In this 
model, the total emission from the pulsar would be due to the propeller 
wind, MDR, and thermal emission from the accretion disk. Wang \& Robertson   
(1985) calculated the angle--integrated thermal bremsstrahlung emissivity 
$j_{B}$ from the heated plasma in a propeller flow. Using their scalings, 
the x--ray luminosity and temperature are given by $L_{x}\sim 4\pi R_{m}^{2}
\delta j_{B}\sim 4.0\times 10^{33}{B_{12}}^{1/2}{\dot{m}_{15}}^{3/4}$ erg 
s$^{-1}$ and $kT\sim 50{\dot{m}_{15}}^{1/2}{B_{12}}^{1/3}$ keV, respectively, 
where $\delta$ is the width of the magnetospheric boundary layer where the 
plasma is heated by the magnetic field. This assumes spherical symmetry, 
so the luminosity will probably be $\sim 10^{33}$ ergs s$^{-1}$ or less 
for the case of a disk geometry. If PSR 1757--24 is not a propeller source, 
we would expect the non-thermal x--ray emission which is characteristic of 
young rotation--powered pulsars (\cite{seward88}) -- e.g. a power law 
emission spectrum with a photon index $\Gamma\sim 2$ and an x--ray 
luminosity given by the empirical relation $L_{x}\propto {(\Omega
\dot{\Omega})}^{1.39}\sim 1.2\times 10^{34}$ erg s$^{-1}$. At a 
distance of $5$ kpc (\cite{gaensler00}), the flux from the non-thermal, 
non-propeller emission would be $4\times 10^{-12}$ ergs cm$^{-2}$ 
s$^{-1}$, which would be easily detectable by {\it XMM} or {\it Chandra}. 
The detection of dimmer, thermal x--ray emission instead of the brighter 
non-thermal emission would be evidence in support of the propeller model. 

Cooler disk blackbody emission could also be visible at optical and 
infrared wavelengths from the accretion disk. The spectrum of propeller 
disks depend on $B$, $\dot{m}$, and the disk orientation with respect 
to the line of sight (e.g. \cite{perna00}), but we can place an upper 
limit on the optical emission from a PSR B1757--24 disk in the 
following manner. Assuming that the majority of the optical disk 
emission originates at the innermost disk radii (\cite{perna00}), 
the upper limit on the optical luminosity is $L_{o}<GM_{\ast}\dot{m}
/R_{m}\sim 0.11 {\dot{m}_{15}}^{9/7}{B_{12}}^{-4/7}$ $L_{\odot}$. 
This estimate ignores heating of the disk by irradiation from 
the pulsar, which dominates the heating only for pulsar 
luminosities $L_{x}>10^{34}$ ergs s$^{-1}$ (\cite{perna00}). 
Assuming $B_{12}=1$, $\dot{m}=1$, and an apparent visual magnitude 
of $-26.78$ for the Sun (\cite{lang80}), the apparent magnitude of 
the disk at $5$ kpc (uncorrected for extinction) is $m_{v}>20.7$. The 
extinction $A_{V}$ can be estimated from the formula $N_{H}=1.79\times 
10^{21}A_{V}$ mag cm$^{-2}$ (\cite{predehl95}), where $N_{H}$ is the 
$HI$ column density along the line of sight. At a distance of $5$ kpc, 
$N_{H}\sim 10^{22}$ cm$^{-2}$ in the Galactic Plane (as can be seen 
from the AXP spectral data in \cite{perna00}), which yields an 
extinction of $A_{V}\sim 5.6$ mag and a lower limit on the disk 
magnitude of $m_{V}>26$. This is comparable to the estimated 
magnitudes of propeller disks around AXPs as calculated by Perna 
et al. (2000).  

If PSR B1757--24 is indeed surrounded by an accretion disk, then fossil 
accretion disks and propeller spin-down may be present, or may have 
been present at one time, in a significant fraction of isolated pulsars. 
This would affect the distributions of pulsar ages, magnetic field 
strengths, and space velocities inferred using the pulsar's $P$ and 
$\dot{P}$ values and the MDR formulae discussed in $\S 1$. Magnetic 
field strengths estimated using Equation (\ref{eq0}), for example, 
would overestimate the true field strengths for both pulsars currently 
undergoing propeller spin-down {\it and} for pulsars which had 
experienced some propeller spin-down in the past (and whose disks 
had subsequently dissipated). The effect on the distribution of radio 
pulsar ages -- and hence the distribution of pulsar velocities 
inferred from their angular positions (e.g. \cite{cordes98}) -- 
depends on whether each pulsar is currently undergoing propeller 
spin-down or not. If a pulsar is currently undergoing propeller 
spin-down, then the MDR timing age $\tau_{MDR}$ is an {\it underestimate} 
of the true age (as exemplified by PSR B1757--24). If a pulsar had 
a propeller disk which then dissipated, its present--day MDR age 
would be an {\it overestimate} of its true age, since the pulsar 
had undergone a period of increased spin-down rate (over the MDR 
spin-down rate) in the past. The latter scenario is probably more 
prevalent in the observed population of radio pulsars, since radio 
pulsations may be scattered and quenched in neutron stars with strong 
propeller flows (e.g. \cite{alpar00}; \cite{fender96}), and therefore 
one might expect a systematic overestimation of radio pulsar ages due 
to fossil disks and propeller spin-down. Curiously, such a systematic 
overestimation of pulsar ages using MDR has been inferred from some 
pulsar population studies (\cite{cordes98}).
 
\section{Summary}

We have shown that the addition of torques from an accretion disk can 
explain the discrepancy between the MDR timing age of PSR B1757--24 
and its true age. This model can be tested through x--ray and optical 
observations of this pulsar. The accretion disk model for pulsar 
B1757--24 leaves open the question of whether or not pulsar B1757--24 
is an unusual and rare object or if it instead reflects a generic 
feature in the evolution of neutron stars. If the former is true, 
the accretion model removes the need (\cite{gaensler00}) to revise 
our current understanding of the physics and astrophysics of neutron 
stars because of this single pulsar. If the latter is true, however, 
the distributions of pulsar magnetic field strengths, ages, and 
velocities will have to be reconsidered to take into account the 
effects of increased spin-down due to accretion disk torques. Observations 
of young pulsars associated with supernova remnants may hold the key 
towards resolving this question, because the pulsar ages can be 
constrained independently of the pulsar temporal parameters. 

\acknowledgements

We thank the referee for helpful comments. This work was performed while 
one of the authors (DM) held a National Research Council-GSFC Research 
Associateship. RER acknowledges support by NASA contract NAS5-30720, 
and REL support from the Astrophysical Theory Program.

{}

\newpage

\begin{figure}
\begin{center}
\plotone{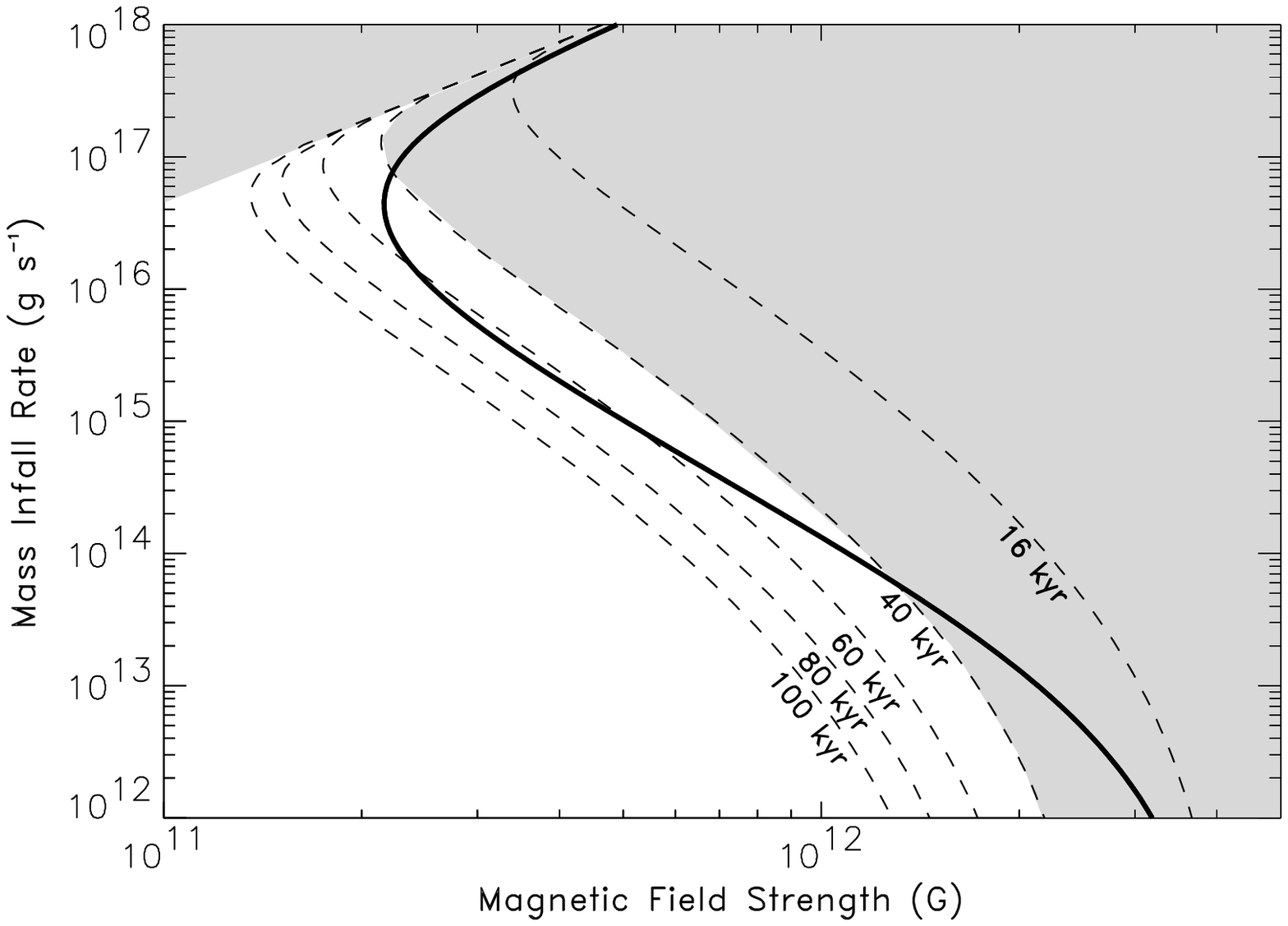}
\end{center}
\caption{~The discrepancy between the magnetic dipole radiation 
(MDR) age of $16$ kyr and the proper motion age of $>39$ kyr for 
PSR B1757--24 may be resolved by the addition of ``propeller'' 
torque due to an accretion disk. This is shown by the plot of 
the calculated age (dashed lines) of pulsar B1757--24 versus the 
neutron star magnetic field strength $B$ and mass infall rate $\dot{m}$, 
assuming a combined spin-down torque due to both magnetic dipole radiation 
(MDR) and an accretion disk formed from supernova debris. The allowed 
combinations of $B$ and $\dot{m}$ fall on portions of the heavy solid 
line (corresponding to the observed period and spin-down rate of the 
pulsar) lying outside the shaded areas excluded by upper limits on the 
pulsar proper motion (Gaensler \& Frail 2000), and the condition 
necessary for the propeller effect ($R_{M}>R_{c}$).}
\end{figure}

\end{document}